\documentclass[traditabstract,longauth]{aa}
\usepackage{hyperref}
\usepackage{natbib}
\usepackage{graphicx}
\usepackage{txfonts}
\usepackage{xspace}
\usepackage{lineno}

\hypersetup{
  colorlinks=true,
  citecolor=blue,
  linkcolor=blue
}

\newcommand{\pks}{\object{PKS\,2155$-$304}}

\begin{document}

\title{Long-term monitoring of PKS\,2155$-$304 with ATOM and H.E.S.S.: investigation of optical/$\gamma$-ray correlations in different spectral states}

\authorrunning{H.E.S.S. Collaboration}
\titlerunning{Long-term monitoring of PKS\,2155$-$304.}

\author{H.E.S.S. Collaboration
\and A.~Abramowski \inst{1}
\and F.~Aharonian \inst{2,3,4}
\and F.~Ait Benkhali \inst{2}
\and A.G.~Akhperjanian \inst{5,4}
\and E.O.~Ang\"uner \inst{6}
\and M.~Backes \inst{7}
\and S.~Balenderan \inst{8}
\and A.~Balzer \inst{9}
\and A.~Barnacka \inst{10,11}
% ...
}

\institute{
Universit\"at Hamburg, Institut f\"ur Experimentalphysik, Luruper Chaussee 149, D 22761 Hamburg, Germany \and
Max-Planck-Institut f\"ur Kernphysik, P.O. Box 103980, D 69029 Heidelberg, Germany \and
Dublin Institute for Advanced Studies, 31 Fitzwilliam Place, Dublin 2, Ireland \and
National Academy of Sciences of the Republic of Armenia,  Marshall Baghramian Avenue, 24, 0019 Yerevan, Republic of Armenia  \and
Yerevan Physics Institute, 2 Alikhanian Brothers St., 375036 Yerevan, Armenia \and
Institut f\"ur Physik, Humboldt-Universit\"at zu Berlin, Newtonstr. 15, D 12489 Berlin, Germany \and
University of Namibia, Department of Physics, Private Bag 13301, Windhoek, Namibia \and
University of Durham, Department of Physics, South Road, Durham DH1 3LE, U.K. \and
GRAPPA, Anton Pannekoek Institute for Astronomy, University of Amsterdam,  Science Park 904, 1098 XH Amsterdam, The Netherlands \and
Obserwatorium Astronomiczne, Uniwersytet Jagiello{\'n}ski, ul. Orla 171, 30-244 Krak{\'o}w, Poland \and
now at Harvard-Smithsonian Center for Astrophysics,  60 Garden St, MS-20, Cambridge, MA 02138, USA
% ...
}

\abstract 
{In this paper we report on the analysis of all the available optical
  and very high-energy $\gamma$-ray ($>$200\,GeV) data for the BL Lac object \pks,
  collected simultaneously with the ATOM and H.E.S.S. telescopes from
  2007 until 2009. This study also includes X-ray (RXTE, {\it Swift})
  and high-energy $\gamma$-ray ({\it Fermi}-LAT) data. During the
  period analysed, the source was transitioning from its flaring to quiescent optical states,
  and was characterized by only moderate flux changes at different wavelengths on the 
  timescales of days and months. A flattening of the optical continuum with an
  increasing optical flux can be noted in the collected dataset,  but
  only occasionally and only at higher flux levels. We did not find any universal
  relation between the very high-energy $\gamma$-ray and optical flux
  changes on the timescales from days and weeks up to several years. 
  On the other hand, we noted that at higher flux levels the source can follow 
  two distinct tracks in the optical flux--colour diagrams, which
  seem to be related to distinct $\gamma$-ray states of the blazar. 
 The obtained results therefore indicate a complex scaling between the optical 
 and $\gamma$-ray emission of \pks, with different correlation patterns holding 
 at different epochs, and a $\gamma$-ray flux depending on the combination 
 of an optical flux and colour rather than a flux alone.} 

\keywords{Radiation mechanisms: non-thermal --- Galaxies: active --- BL Lacertae objects: PKS\,2155$-$304 --- Galaxies: jets --- Gamma rays: galaxies}

\maketitle

\section{Introduction}

Blazars are those active galactic nuclei (AGNs) for which the observed radiative output
is dominated by a broadband non-thermal emission of a relativistic jet pointing at a very 
small angle to the line of sight \citep[e.g.][]{begelman84,urry95}. This emission, strongly
Doppler-boosted in the observer's rest frame, extends from radio up to high-energy (HE) and very
high-energy (VHE) $\gamma$ rays, and is strongly variable on different timescales from years 
down to minutes \citep[e.g.][]{wagner}. The spectral energy distribution (SED) of blazars is characterized by two
distinct components manifesting as low- and high-energy peaks in the $\nu - \nu F_{\nu}$
representation (where $\nu$ is the frequency and $F_{\nu}$ is the corresponding flux). The established model for the broadband blazar emission ascribes the
low-energy spectral component to the synchrotron radiation of ultra-relativistic electrons 
accelerated within the inner parts of a magnetized jet (sub-parsec and
parsec scales). The high-energy component is widely believed to be due to the inverse-Compton (IC) emission of the same electron 
population, involving either jet synchrotron photons as seeds for the IC scattering 
\citep[synchrotron self-Compton model, SSC; e.g. ][]{konigl, marscher}, or various photon 
fields originating outside of a jet \citep[external-Compton models; e.g. ][]{dermer92, 
sikora94}. The emerging picture is that the SSC model is the most appropriate for 
blazars of the BL Lacertae type which accrete at lower rates than in other blazars and as such lack intense 
circumnuclear photon fields, while the external-Compton scenarios are best applied in 
modelling quasar-hosted blazars known for their high accretion rates and rich 
circumnuclear environment \citep[see e.g.][]{ghisellini10}.
The high-energy peak can be alternatively modelled in the framework of 
hadronic scenarios \citep[see e.g.][]{boe07}.
The frequency of the synchrotron peak in blazars spectrum defines the classification for low-, intermediate- and high-frequency-peaked objects (LBL, IBL, and HBL, respectively) \citep[see, e.g.][]{padovani95,fossati98}.

The source\pks\ (redshift $z = 0.117$), classified as a HBL object is one of the brightest 
blazars in the southern hemisphere. 
It was discovered in the radio frequencies as part of the Parkes survey
\citep{Shimmins74}, and identified as a BL Lac-type source by \cite{Hewitt80}.
Since then it has been the target of a number of X-ray monitoring
programmes. These studies indicated a red-noise 
type of variability at X-ray frequencies with the characteristic variability timescale 
of the order of days, and only little power corresponding to intraday flux changes 
\citep{Zhang99,Zhang02,Tanihata01,Kataoka01}. Frequent X-ray flares are 
typically characterized by symmetric profiles, superimposed short ($\sim 10$\,ks) 
smaller-amplitude flickering, and a variety of soft and hard lags changing from epoch to epoch. 
The amplitude of the variability correlates with both the flux of the source and the 
photon energy. The X-ray spectral variations are consistent with the ''harder-when-brighter'' 
behaviour \citep{Zhang05,Zhang06a,var4}.

As a particularly bright BL Lac object, \pks\ was also targeted by several extensive optical/UV studies 
\citep{Courvoisier95, Pesce97,Pian97}. These campaigns established the intraday variability nature of the 
source with the shortest flux doubling timescales of about $15$\,min, detected, however, in 
only a few isolated epochs \citep{Paltani97,Heidt97}. The red-noise type of  optical variability, 
comparable to that observed at X-ray frequencies, persists up to variability timescales of 
about 3 years \citep{Kastendieck}. In several epochs a clear bluer-when-brighter evolution of 
the source has been found \citep{Paltani97}, a behaviour  that appears to beuniversal in BL Lac objects 
\cite[see e.g.][]{Carini1992, Ghisellini1997, Raiteri2001, Ikejiri2011}.

A comparison between optical/UV and X-ray temporal and spectral characteristics of \pks\ indicates 
that the peak of its synchrotron continuum is located around UV frequencies, only occasionally shifting 
to longer (optical) wavelengths \citep[see][]{foschini08}. In addition, the variability amplitudes at optical and UV
frequencies are always significantly smaller than those observed at X-ray frequencies. In general, a variety
of optical/UV--X-ray correlation patterns have been found in different epochs and datasets, although several 
authors have noted that such correlations are in general stronger and more pronounced at shorter
variability timescales and during the enhanced activity epochs of the source: as the flux decreases and 
variability timescales get longer, the optical/UV--X-ray correlations become weaker, and at the same time 
the lags between flux changes at optical/UV and X-ray frequencies increase \citep[see e.g.][]{brinkmann,
Urry97,Zhang06b,Dominici04,Osterman07}.

High-energy and very high-energy $\gamma$-ray emission of \pks\ was discovered with EGRET 
\citep{vestrand} in the energy range from 30\,MeV to 10\,GeV, and with the University of Durham Mark 6 
telescope above 300\,GeV \citep{Chadwick99}, respectively. The first
detection with the High Energy Stereoscopic System (H.E.S.S.) of the source 
was made during the July and October 2002 observations \citep{hess2004}. The first multiwavelength 
(MWL) campaign targeting the blazar and organized by the H.E.S.S. Collaboration was conducted 
in 2003 \citep{hess2005}. The source, found in its low state, was monitored simultaneously in four 
different bands: in X-rays with RXTE/PCA, in the optical range with
ROTSE, at radio frequencies with the 
Nancay Radio Telescope, and in VHE $\gamma$ rays with H.E.S.S. (with an 
incomplete array). No clear correlations 
between H.E.S.S. and lower-energy fluxes were found at that time. From 2004 until now \pks\ was frequently 
observed with the four H.E.S.S. telescopes, and this resulted in the detection of two exceptional $\gamma$-ray 
flares in July 2006, with the outburst flux 40 times the average flux, and  flux doubling timescales in the
VHE regime of the order of a few minutes \citep{var1}. The MWL data collected around the time of the 
2006 flares were presented and extensively discussed in \citet{var2} and \citet{var4}. These revealed a  strong
X-ray--VHE flux correlation at high flux levels, weakening however at lower flux levels, and characterized 
in addition by much smaller X-ray flux changes (factor of $\sim 2$) than those observed in the VHE range. 
No universal optical--VHE correlations were found in the 2006 flaring data.

Detailed statistical analysis of all the VHE data collected for \pks\ with H.E.S.S. during the period 2005-2007 
was presented in \citet{var3}. The results of the analysis revealed again the red-noise type of the flux changes
albeit with relatively short characteristic variability timescales $\lesssim 1$\,day \citep[in particular during the 2006 flaring
period; see also][]{var1}, the variability amplitude characterized by fractional r.m.s. correlated with the photon energy, and the excess r.m.s. correlated with the flux. The pronounced spectral changes in the VHE regime were found to be different for the 
low- and high-flux levels.

Between August 25 and September 6,2008 \pks\ was observed simultaneously with H.E.S.S., {\it Fermi}-LAT, 
RXTE and ATOM \citep{varLAT}. The source was found again in its low state, with the average $>200$\,GeV 
flux about $15\%$ of the flux from the Crab Nebula. The `night-by-night comparison' of the VHE and optical fluxes at that 
time (11 days of observations) indicated, for the very first time in the case of this target, some 
hints of a positive correlation (Pearson's correlation coefficient in the range 0.77-0.86 with uncertainties $<0.09$).
In support of the observed trend, the performed broadband SSC modelling confirmed for the broadly expected 
parameters of the source characterizing its quiescent periods, that these are indeed the 
electrons emitting synchrotron photons at optical frequencies which produce the bulk of the observed $\gamma$ rays 
via the SSC process. These findings prompted us to analyse all the available optical and VHE data for the source 
collected simultaneously in 2007-2009 with the ATOM and H.E.S.S. telescopes, respectively, and in addition the X-ray 
and high-energy $\gamma$-ray data. Except for the short time 
interval between August 25 and September 6, 2008 discussed in \citet{varLAT}, the MWL data presented here have not been 
published before. 

\section{Observations and  data analysis}

During the last years, \pks\ has been observed multiple times at different frequencies. In this paper we
analyse the 2007-2009 data collected with H.E.S.S. in VHE $\gamma$-rays (see Sect.,\ref{hess_data}), {\it Fermi}-LAT
in HE $\gamma$ rays (Sect.,\ref{fermi_data}), RXTE in X-rays (Sect.,\ref{rxte_data}), and the
ATOM telescope in the optical band (Sect.,\ref{atom_data}).

\begin{figure}[t]
\centering{\includegraphics[width=0.48\textwidth]{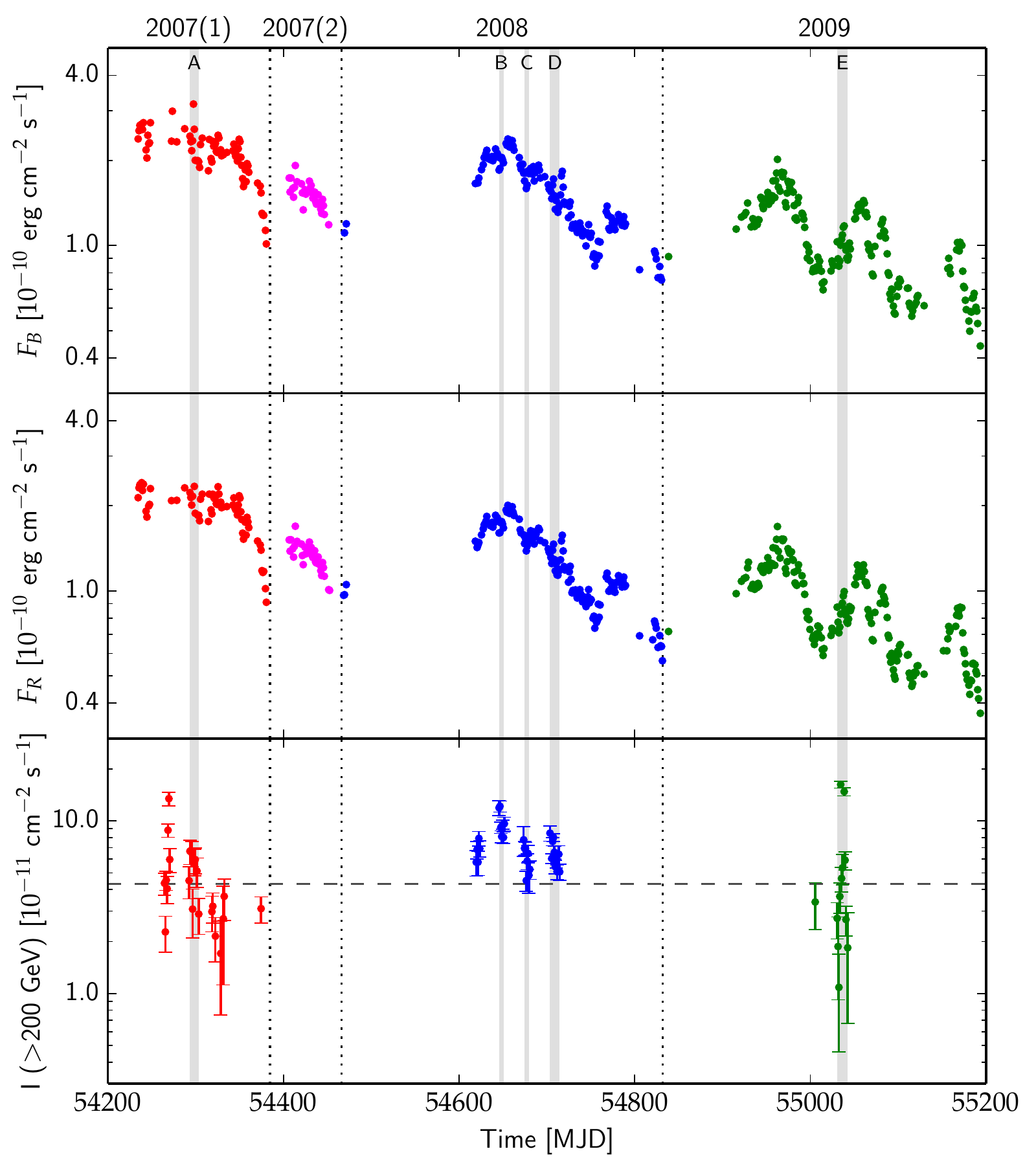}}
\caption[]{Optical (ATOM; $B$ and $R$ bands) and VHE $\gamma$-ray (H.E.S.S.) lightcurves of \pks\ during the period 2007--2009 (upper, middle, and bottom panels, respectively). The observations performed in 2007 until MJD54385 are denoted by red symbols, in 2007 after MJD54385 by magenta symbols, in 2008 by blue symbols, and in 2009 by green symbols; these epochs are also separated by the dotted vertical lines. For the two upper panels the error bars are within the size of the points. The optical and VHE lightcurves are binned in one-night intervals. We note uneven sampling of the lightcurves at optical and VHE frequencies, as well as different integration times between the optical and VHE exposures.
The horizontal line in the lower panel indicates the quiescent state
of the source, defined in \cite{var3}. The vertical grey lines
indicate the A-E intervals defined in Sect.\ref{opt-gamma-cor}.
}
\label{atom_lc}
\end{figure}

\subsection{VHE $\gamma$-ray data from H.E.S.S.}
\label{hess_data}

The  H.E.S.S. telescopes are designed to observe $\gamma$ rays 
with energies between 100\,GeV and 100\,TeV, 
located in Namibia, near the Gamsberg mountain
\citep[see][]{crab}. 
 Until 2012 the instrument consisted of four Imaging Atmospheric Cherenkov telescopes, now
 it is an array of five telescopes. 

The blazar\pks\ has been observed frequently with H.E.S.S. during each visibility window (May to December) since 2004, 
being routinely detected on a nightly basis. The data analysed here correspond to all 
the available observations collected from 2007 to 2009 by an array
configuration with three or four telescopes. The data were selected using standard data-quality selection criteria, and this resulted in 12 nights of 
observations in 2007 (MJD54293-54375), 25 nights in 2008 (MJD54620-54715), and 12 nights in 2009
(MJD55005-55043). We note that the night exposures of 
\pks\ with H.E.S.S. consist typically of one or a few observational
runs, each about 28 minutes long.
The data were calibrated using the standard H.E.S.S. calibration method; 
loose cuts were used for event selection \citep{crab} and the background estimated using the reflected 
background method \citep{Berge}. The standard Hillas analysis \citep{crab} was applied,
and the VHE spectra for the different epochs considered were derived using a forward-folding 
maximum likelihood method \citep{piron}.
Similar results to those presented below (Sect.,\ref{sec-result}) were also
obtained using an alternative analysis method  \citep{Becherini} and an independent calibration chain.

\subsection{HE $\gamma$-ray data from {\it Fermi}-LAT} 
\label{fermi_data}

The \emph{Fermi} Gamma-ray Space Telescope was launched in 2008 June 11. 
The Large Area Telescope (LAT) onboard the \emph{Fermi} satellite 
is a pair conversion telescope sensitive to photons in a very broad energy band from $\sim 20$\,MeV 
to $\gtrsim 500$\,GeV, uniformly scanning  the entire sky every three hours \citep{atwood}. 

The LAT observations analysed in this paper include all the data taken
from the mission start to MJD54716 (for 2008), as well as the more recent  data simultaneous with and fully overlapping the H.E.S.S. exposure of the target in 2009. The LAT Science Tools version \texttt{v9r31} with 
the response function \texttt{P7SOURCE\_V6} were used for the LAT data reduction. 
For the spectral analysis only the events with zenith angle less than $105^{\circ}$ and in the energy 
range from 100\,MeV up to 100\,GeV were selected, with $15^{\circ}$ region of interest (ROI) centred on
\pks. The binned maximum likelihood method \citep{Mattox96} was applied in the analysis.
The maximum zenith angle cut is chosen to exclude time periods, when part of the region of interest is close to the Earth's Limb, resulting in elevated background levels. 
The Limb gamma-ray production by cosmic ray interactions in the Earth's atmosphere is discussed by \citet{Fermi_limb}.
The Galactic diffuse background was modelled using the \texttt{gll\_iem\_v05} map
cube and the extragalactic diffuse and residual instrument backgrounds were
All the sources from the {\it Fermi}-LAT Second Source Catalog 
\citep[2FGL,][]{2fgl} inside the ROI of \pks\ were modelled.
The sources from 2FGL, which were outside the ROI but  bright enough to have an impact on the analysis have also been modelled.

\subsection{X-ray data from RXTE} 
\label{rxte_data}
In 2008 and 2009 \pks\ was observed in X-ray range with the Proportional Counter Array (PCA) detector onboard the Rossi X-ray Timing 
Explorer (RXTE) satellite \citep{bradt}. The 2008 observations discussed here were 
presented previously in \cite{varLAT}. The unpublished data were collected in  July 28 and 31, 2009. 
For the RXTE/PCA data analysis \texttt{HEASOFTv6.12} software package was used. 

\subsection{Optical data from ATOM} 
\label{atom_data}

Active Galactic Nuclei targeted with H.E.S.S. are monitored with the 75\,cm Automatic
Telescope for Optical Monitoring (ATOM) located in Namibia
at the H.E.S.S. site. The telescope has been operational since November 2006, and is
conducting observations in four different filters: $B$ (440\,nm), $V$ (550\,nm), $R$ (640\,nm) 
and $I$ (790\,nm). The detailed description of the instrument can be found in \cite{Hauser}.

The ATOM data analysed in this paper correspond to integration times between 60\,s and 200\,s
for each daily exposure. 
The photometric flux scale was calibrated using reference stars from the Landessternwarte (Universit{\"a}t Heidelberg, K{\"o}nigstuhl) AGN finding charts\footnote{http://www.lsw.uni-heidelberg.de/projects/extragalactic/charts}; the 
photometric accuracy achieved is between $0.01$\,mag and $0.02$\,mag
for $B$ and $R$ filters. 
In addition to the standard automatic analysis of the collected data, all the raw images were also checked 
manually, and data taken during bad weather conditions were rejected from the final dataset. 
The observed magnitudes have been corrected against the Galactic extinction
based on the model from \cite{Schlegel98} with the most recent recalibration by \cite{Schlafly}, 
using the NED Extinction Calculator ($A_B = 0.078$\,mag, $A_R = 0.047$\,mag). 
In the analysis presented below, the data are also corrected for the contribution of the host galaxy, using  
the template of an elliptical galaxy provided by \cite{Fukugita95} and observations in Gunn filter $i$ obtained by \cite{Falomo91},
with an assumed de Vaucouleurs profile of the starlight. 
The measured brightness of \pks\ was transformed to the ATOM aperture of $4''$ using Eq. (4) in \cite{Young76}.

\begin{figure}[t]
\centering{\includegraphics[width=0.48\textwidth]{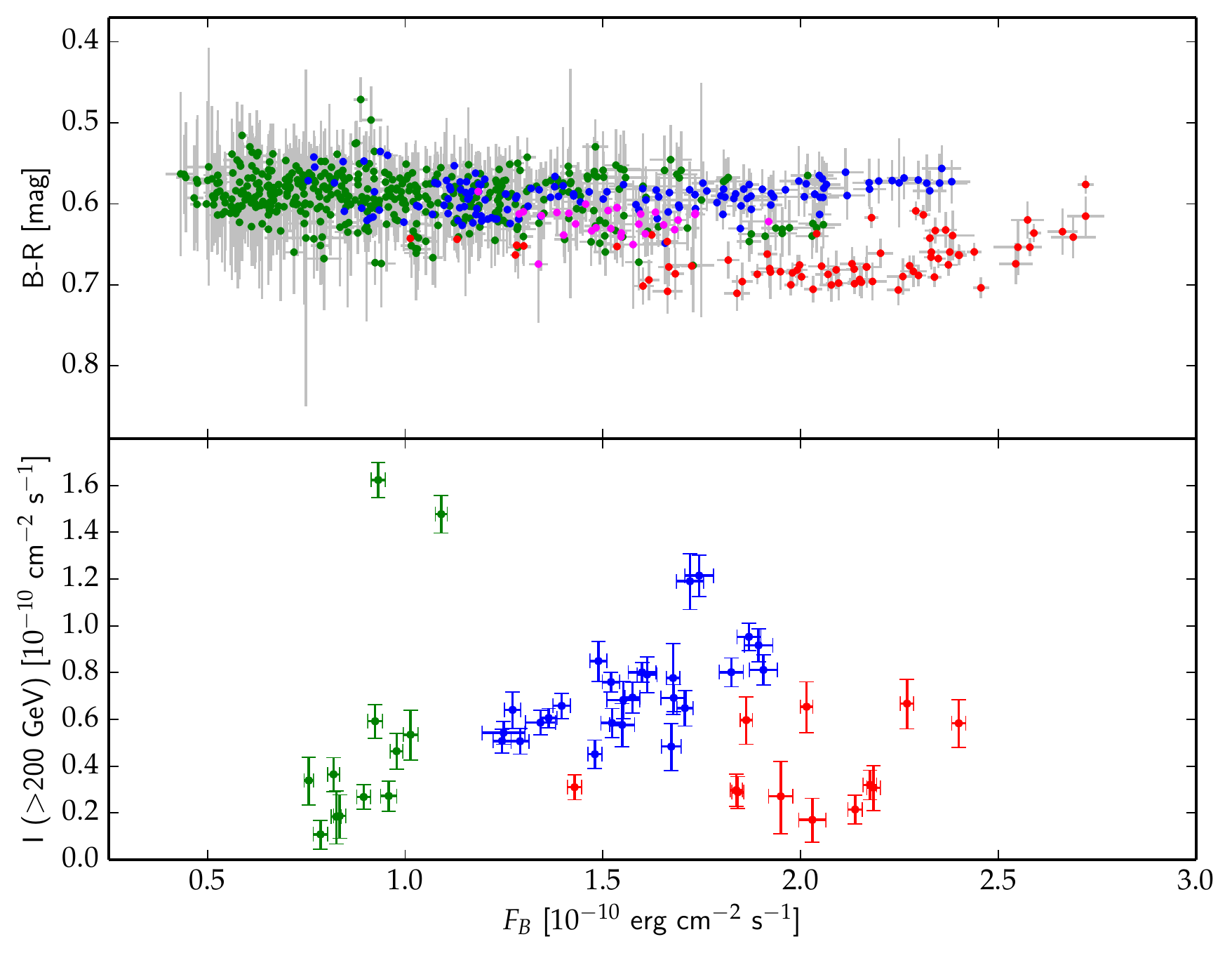}}
\caption{The upper panel shows the $B-R$ colour vs. the $B$-band energy flux for \pks\ during the analysed period 2007-2009. The lower panel presents the corresponding VHE $\gamma$-ray photon flux of the source also as a function of the $B$-band energy flux. The applied colour coding is the same as in Figure~\ref{atom_lc}. We note two distinct optical spectral states appearing at high flux levels in the upper panel, and the correspondingly distinct optical--VHE correlation pattern emerging in the lower panel (green and blue points), for details see $Sect.;$\ref{opt-gamma-cor}. The colour uncertainties are calculated as r.m.s.}
 \label{hess_b}
\end{figure}

\begin{table*}
\centering
\begin{tabular}{l|c|c|c|c|c|c}
\hline
\hline
Year & $F_{var}(B)$ & $F_{var}(R)$ & $\tau_{d}^{(i)}$ & $\tau_{d}^{(ii)}$ & $\langle [\nu F_{\nu}]_B \rangle$ & $\langle [\nu F_{\nu}]_R \rangle$\\
 & & & [day] & [day] & [$\times 10^{-10}$\,erg\,cm$^{-2}$s$^{-1}$] & [$\times 10^{-10}$\,erg\,cm$^{-2}$s$^{-1}$] \\
\hline
$2007^{(1)}$  & 0.188 $\pm$ 0.015  & 0.166 $\pm$ 0.015& 3.37 $\pm$ 0.10 &
3.97 $\pm$ 0.17 & 2.12 $\pm$ 0.40 & 1.93 $\pm$ 0.33 \\
$2007^{(2)}$ & 0.117 $\pm$ 0.016 & 0.127 $\pm$ 0.017 & 5.65 $\pm$ 1.00 & 7.63  $\pm$ 0.57
& 1.50 $\pm$ 0.18 & 1.30 $\pm$ 0.17 \\
$2008$ & 0.288 $\pm$ 0.018 & 0.278 $\pm$ 0.017 & 6.48 $\pm$ 0.64 & 7.51 $\pm$ 0.28
& 1.51 $\pm$ 0.44 & 1.28 $\pm$ 0.37 \\
$2009$ & 0.327 $\pm$ 0.018 & 0.308 $\pm$ 0.016 & 4.43 $\pm$ 0.46 & 5.16 $\pm$ 0.36
& 1.05 $\pm$ 0.34 & 0.88 $\pm$ 0.29 \\
\hline
\end{tabular}
\caption[]{The evaluated fractional r.m.s. (Eq. \ref{equ_fvar}), doubling timescales (Eq. \ref{equ_tau}), and average energy fluxes in $B$ and $R$ filters for \pks\ in different epochs during the period 2007-2009.
The meaning of $\tau_d^{(i)}$ and $\tau_d^{(ii)}$ can be found in $Sect.;$\ref{opt-var-char}. }
  \label{table_1}
\end{table*}

\section{Results}
\label{sec-result}

\subsection{VHE and optical lightcurves}

The three-year (2007-2009) lightcurves of \pks\ in $B$ and $R$ filters are
shown in Figure~\ref{atom_lc} (upper and middle panels). The colour coding in the figure
corresponds to the different years of observations and these are the epochs
corresponding roughly to the subsequent H.E.S.S. visibility windows. In the
case of the 2007 data, we divide them further into 2007$^{(1)}$ and
2007$^{(2)}$ sub-epochs (denoted by red and magenta symbols, respectively),
with the borderline date MJD54385, as justified by the results of the analysis
presented below.  We note that there were no H.E.S.S. data taken during
2007$^{(2)}$.  The average energy fluxes in both filters over the analysed
intervals are given in Table~\ref{table_1}.

The VHE $\gamma$-ray lightcurve of the source for the same time period is shown in the 
lower panel of Figure~\ref{atom_lc}. Each point in the presented lightcurves corresponds to one night 
of observation, but integration times differ between the optical and VHE exposures. 

\subsection{Colour vs. flux diagram}

The diagram illustrating the $B-R$ colour vs. $B$-band energy flux for \pks\ during the analysed 
period 2007-2009 is presented in Figure~\ref{hess_b}. As shown, at the higher flux levels two distinct 
branches appear in the diagram: the `upper' branch corresponding predominantly to the 2008 epoch, 
and the `lower' branch consisting exclusively of the data collected during the 2007$^{(1)}$ epoch
and displaying steeper optical spectra (i.e. larger values of the optical spectral index 
$\alpha_{RB}$ defined for the flux spectral density as $\alpha = - \ln F_{\nu} / \ln \nu$). 
Both branches are similarly characterized by a flattening of the optical continuum with increasing fluxes, 
although this `bluer--when--brighter' trend is pronounced for the lower branch only at magnitudes 
$B< 13.0$\,mag, corresponding roughly to the energy fluxes $[\nu F_{\nu}]_B > 2 \times 
10^{-10}$\,erg\,cm$^{-2}$\,s$^{-1}$.
The effect can be quantified by the colour-flux Pearson's correlation coefficients:
we derive $C = 0.16 \pm 0.07$ for the 2008 epoch regardless on the flux level, 
while $C = 0.61 \pm 0.07$ and $C = -0.53 \pm 0.19$ for the 2007$^{(1)}$ epoch 
taking into account the $B<13$\,mag and $B>13$\,mag datapoints, respectively.

Since the distinct lower branch in the colour--flux diagram consists of a single-epoch dataset
only,  care must be taken to check if it is not an artefact of particular weather conditions affecting optical
flux measurements. For this purpose we have carefully investigated the ATOM data taken for the
other blazars (PKS\,2005$-$489 and H\,2356$-$309) during the same nights as the 2007-2009 data 
for \pks. In all the cases various flux and colour changes were observed, for sources other than \pks\ 
the 2007$^{(1)}$ epoch does not show separation in the colour--flux diagrams.
Distinct colour--flux states are also not observed for the comparison star, which is separated on the sky
from \pks\ by $\sim 1$\,arcmin, and which is always present on the same raw images as the 
blazar. One of the comparison stars has been used as a reference to test the reality of this effect.  
Thus, we can exclude atmospheric effects as playing a significant role, and conclude that the two distinct 
optical states manifesting as two branches on the colour-flux diagram at high flux levels are intrinsic 
to the studied object.

\subsection{Optical variability characteristics} 
\label{opt-var-char}

In order to characterize quantitatively the mean variability of \pks\ in the optical band during the
analysed period, we calculate the fractional r.m.s. of the source defined as
\begin{equation}
 F_{var}=\frac{\sqrt{S^2-\sigma^2_{err}}}{\bar{\Phi}} \, ,
 \label{equ_fvar}
\end{equation}
where $\bar{\Phi}$ is the mean flux of $N$ measurements, $\sigma^2_{err}$ is the mean square 
error, and $S^2$ is the variance \citep[e.g.][]{edelson}. The value of  $F_{var}$ is calculated in the $B$ and $R$ bands 
separately for the different epochs considered, and the results are given in  Table~\ref{table_1}.
As shown, the higher-flux 2007$^{(1)}$ epoch is characterized by a smaller value of the fractional
r.m.s. ($F_{var} \lesssim 20\%$) when compared with the dimmer (on average) 2008 and 2009 epochs 
($F_{var} \simeq 30\%$), although the difference is only modest, and as such should not be 
over-emphasized. 
The fractional variability amplitudes derived by \cite{bonning12} in
different optical filters for \pks\ observed with SMARTS
in the years 2008-2010 are in agreement with the values provided here.
We note that the amplitudes of the optical flux changes 
observed here are smaller than those typically observed in \pks\ at X-ray or VHE 
$\gamma$-ray energies, even during quiescent states of the source.

Another parameter characterizing the source variability is the doubling/halving timescale, which 
provides an important constraint on the spatial scale of the dominant emission region through the causality
requirement. This doubling timescale, $\tau_d$, may be evaluated as either (i) the smallest value of the quantity
\begin{equation}
 \tau_{k,m}=\left|\frac{\Phi \, \Delta T}{\Delta
     \Phi}\right|=\left|\frac{(\Phi_k + \Phi_m ) \, (T_k-T_m)}{2 \,
     (\Phi_k-\Phi_m)}\right| ,
\label{equ_tau}
\end{equation}
where $\Phi_j$ is a flux of the source at the time $T_j$ for all pairs of points $(k,m)$ in the lightcurve, or
(ii) the mean of the five smallest $\tau_{k,m}$ within the whole dataset \citep[see e.g.][]{Zhang99}. 
The two corresponding different values of $\tau_d$ in the $B$-band for all the considered epochs are given in 
Table~\ref{table_1}. As shown, the 2007$^{(1)}$ epoch is characterized by marginally shorter flux doubling
timescales than the subsequent epochs.
On the other hand the derived values are  smaller than flux doubling timescales claimed by \cite{chatterjee12} for the source based on the SMARTS 2008-2010 data ($\sim$ 22-33\,days).
It should be stressed that doubling timescales shorter than a day could not be 
investigated in the presented dataset, as the analysed non-uniformly sampled lightcurves of the source 
are binned in one-night intervals and hence, according to Equation~\ref{equ_tau} above, 
$\tau_{k,m}^{(min)} =  (T_k-T_m)/2 \geq 0.5$\,day. We can therefore neither claim nor exclude 
a presence of intra-day optical flux changes in \pks\ during the analysed period.

\subsection{Optical--VHE $\gamma$-ray correlations}
\label{opt-gamma-cor}

The VHE $\gamma$-ray light curve of \pks\ (photon flux $I_{> 200\,{\rm GeV}}$) is shown in the lower panel of
Figure~\ref{atom_lc}, where all the selected H.E.S.S. observation runs are combined to derive 
nightly flux values. The corresponding VHE $\gamma$-ray flux changes as a function of the $B$-band
energy flux of the source are presented in the lower panel of Figure~\ref{hess_b}.
The plot shows only the nights when both optical and VHE $\gamma$-ray data are available.

Both the uneven sampling of the lightcurves at optical and VHE frequencies, as well as different 
integration times between the optical and VHE exposures, complicate the analysis of the 
optical--VHE $\gamma$-ray flux correlations.
We note here again that in \cite{varLAT} a positive Pearson's correlation coefficient 
$0.77 \leq C \leq 0.86$ between the VHE and optical bands was derived for 12 days of 
simultaneous observations. In the case of the entire 2007-2009 dataset analysed in this paper, 
some general positive correlation seems to be present in the lower panel of 
Figure~\ref{hess_b}, albeit only at low optical flux level, as the correlation is 
apparently absent for the higher-flux/steep-spectrum state (`lower branch' in 
Figure~\ref{hess_b} corresponding to the 2007$^{(1)}$ epoch). 

\begin{figure}[]
\centering{\includegraphics[width=0.48\textwidth]{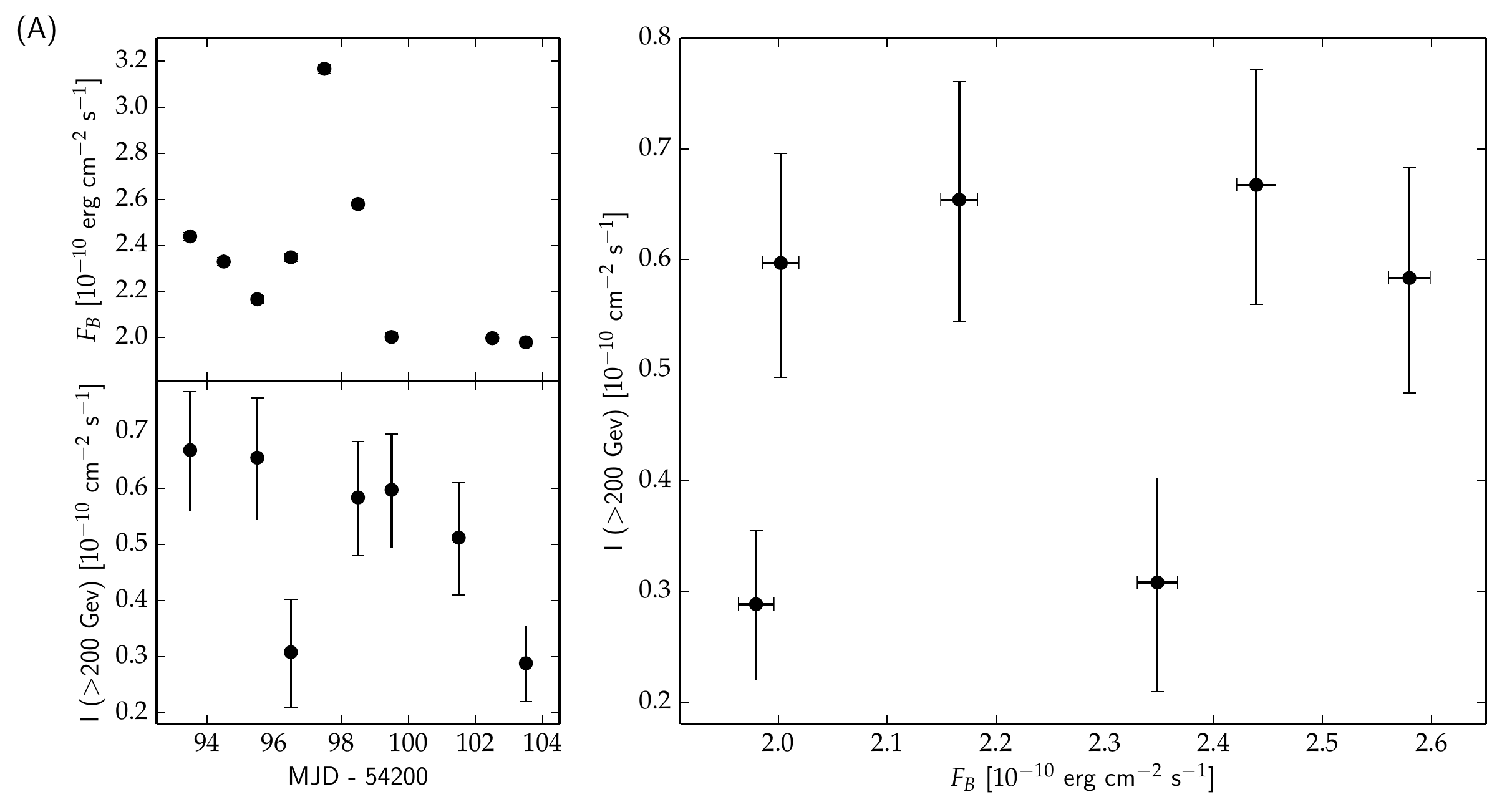}}
\centering{\includegraphics[width=0.48\textwidth]{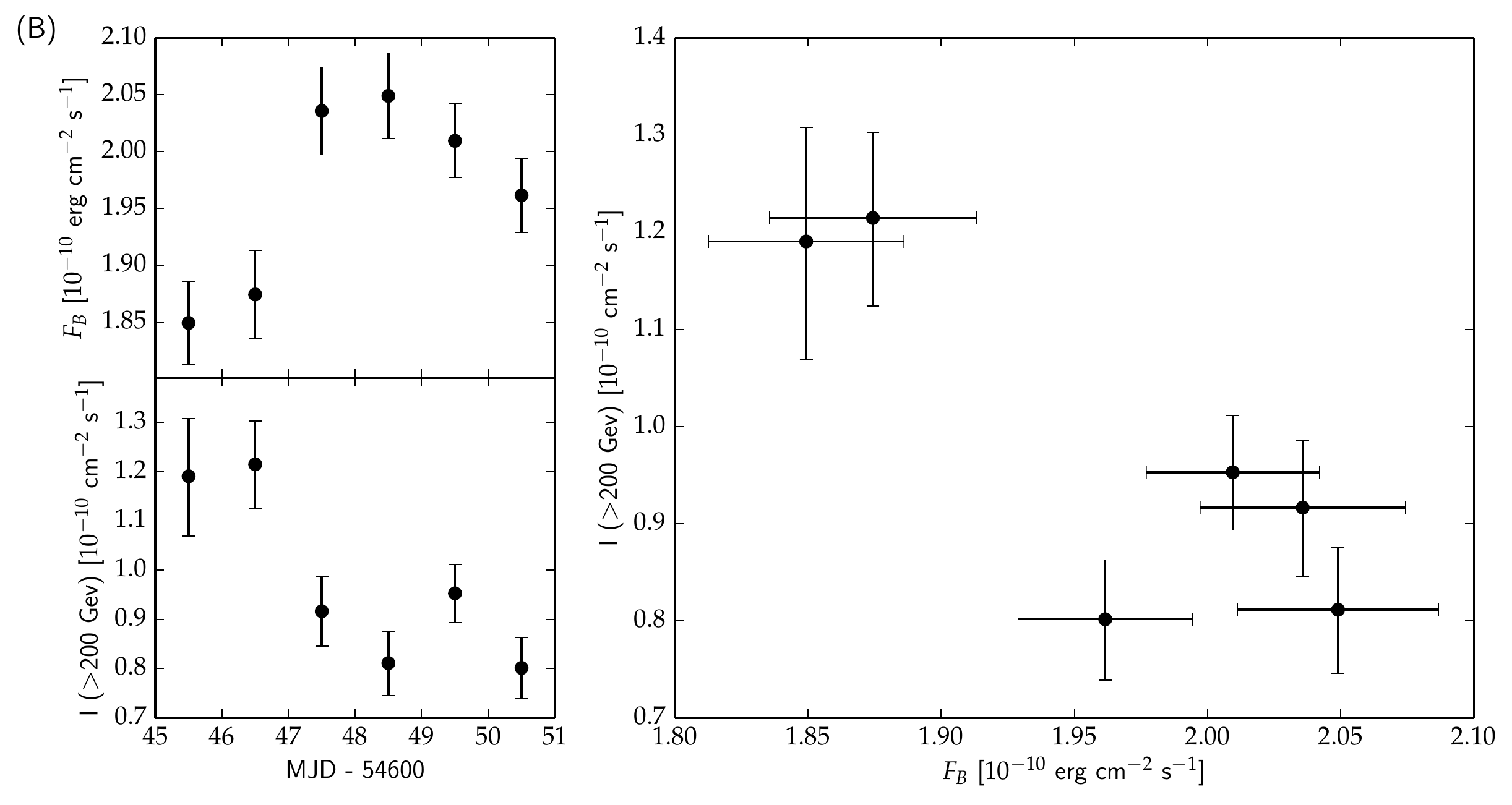}} 
\centering{\includegraphics[width=0.48\textwidth]{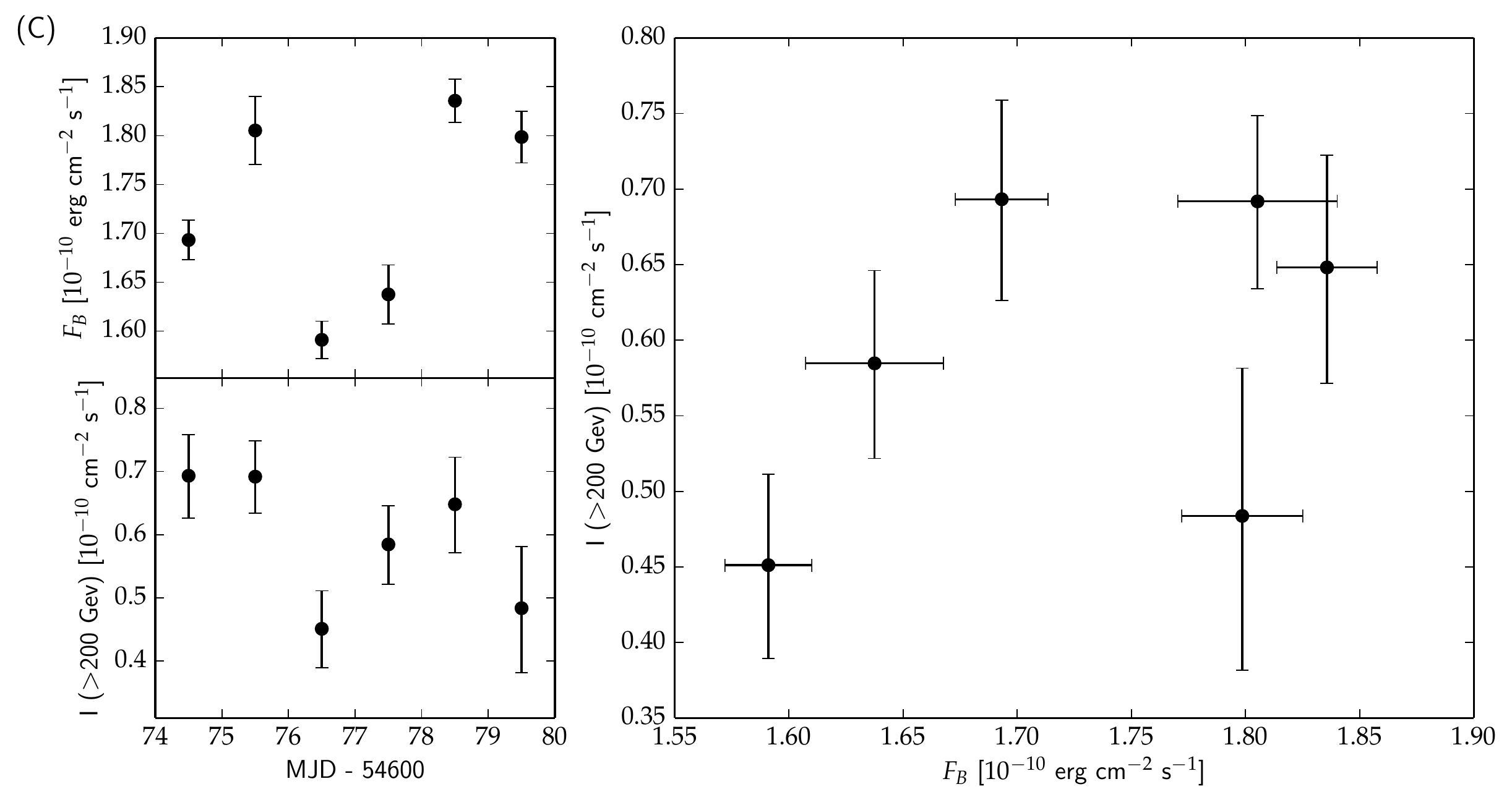}} 
\centering{\includegraphics[width=0.48\textwidth]{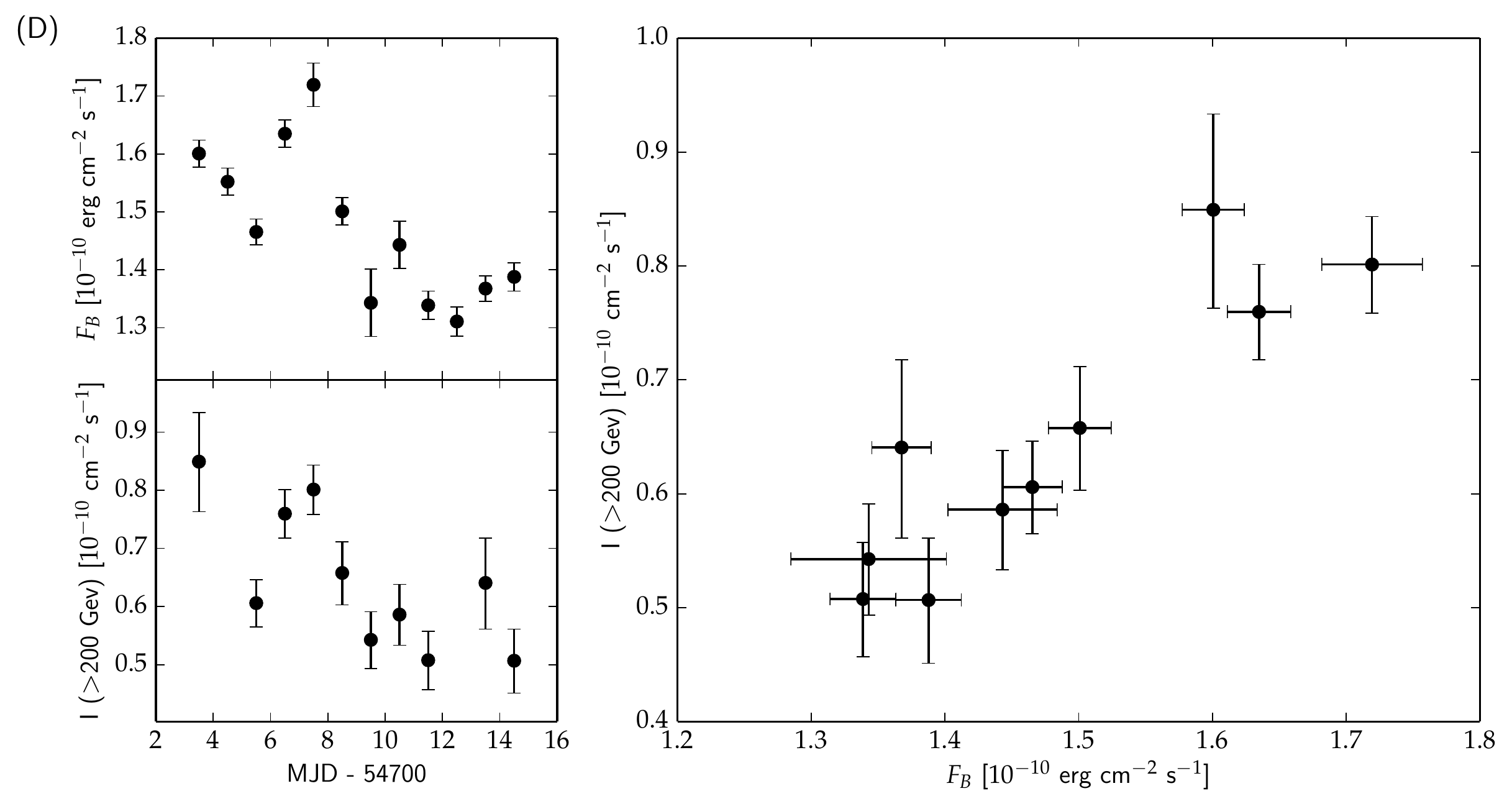}}
\centering{\includegraphics[width=0.48\textwidth]{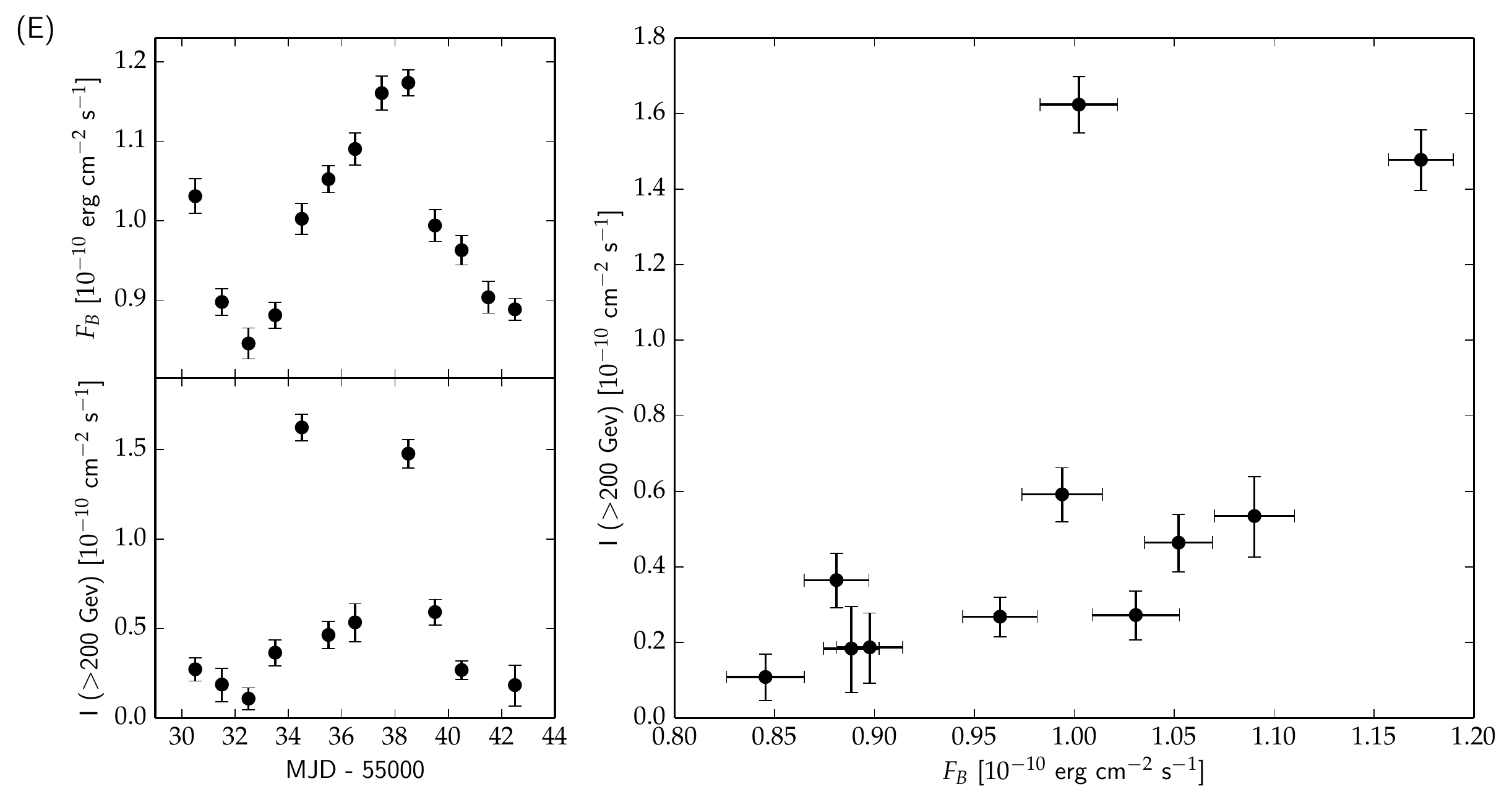}}
\caption[]{ATOM and H.E.S.S. lightcurves and flux-flux diagrams for the intervals A-E (see Table~\ref{table_2}).}
\label{four_corr}
\end{figure}

Within the whole analysed 2007-2009 dataset, in addition to the campaign discussed in \cite{varLAT} 
and denoted below as `interval D', there are four other intervals (hereafter `A-C' and `E') with simultaneous night-by-night 
observations involving more than four consecutive nights (in some cases with one-day gaps). 
These are all presented in different panels of Figure~\ref{four_corr}. The correlation coefficients and correlation 
slopes derived for these intervals are listed in Table~\ref{table_2}. 
As shown in the figure and quantified in the table, there is no universal relation between VHE $\gamma$-rays and the optical flux changes on a weekly timescale: the fluxes are correlated on some occasions, but uncorrelated or even anti-correlated on the other occasions. 
Hence, we conclude that there is not always a positive correlation between the optical and the VHE and that the one reported by \cite{varLAT} was particular to the state of the source at this epoch.

\begin{table*} 
\centering
\begin{tabular}{c|c|c|c|c}
\hline
\hline
 Interval & Time [MJD]  & Correlation coefficient & Correlation slope & Chance probability\\
\hline
A & 54294$-$54304 & 0.29 $\pm$ 0.23 & -- & -- \\
B & 54646$-$54651 & $-$0.84 $\pm$ 0.33 &  $-1.96 \pm 0.61$ & 0.22\\
C & 54675$-$54680 & 0.43 $\pm$ 0.31 & -- & -- \\
D & 54704$-$54715 & 0.89 $\pm$ 0.24 & $0.88 \pm 0.15$ & 0.05\\
E & 55031$-$55043 & 0.63 $\pm$ 0.10 & $3.5 \pm 1.4$ & 0.22\\
\hline
\end{tabular}
\caption[]{VHE/optical flux correlation coefficients, correlations slopes, and chance probabilities derived for the intervals A-E constituting simultaneous night-by-night ATOM and H.E.S.S. observations of \pks\ involving more than four consecutive nights (in some cases with one-night gaps) during the analysed period 2007-2009.}
  \label{table_2}
\end{table*}

\begin{figure}[]
\centering{\includegraphics[width=0.48\textwidth]{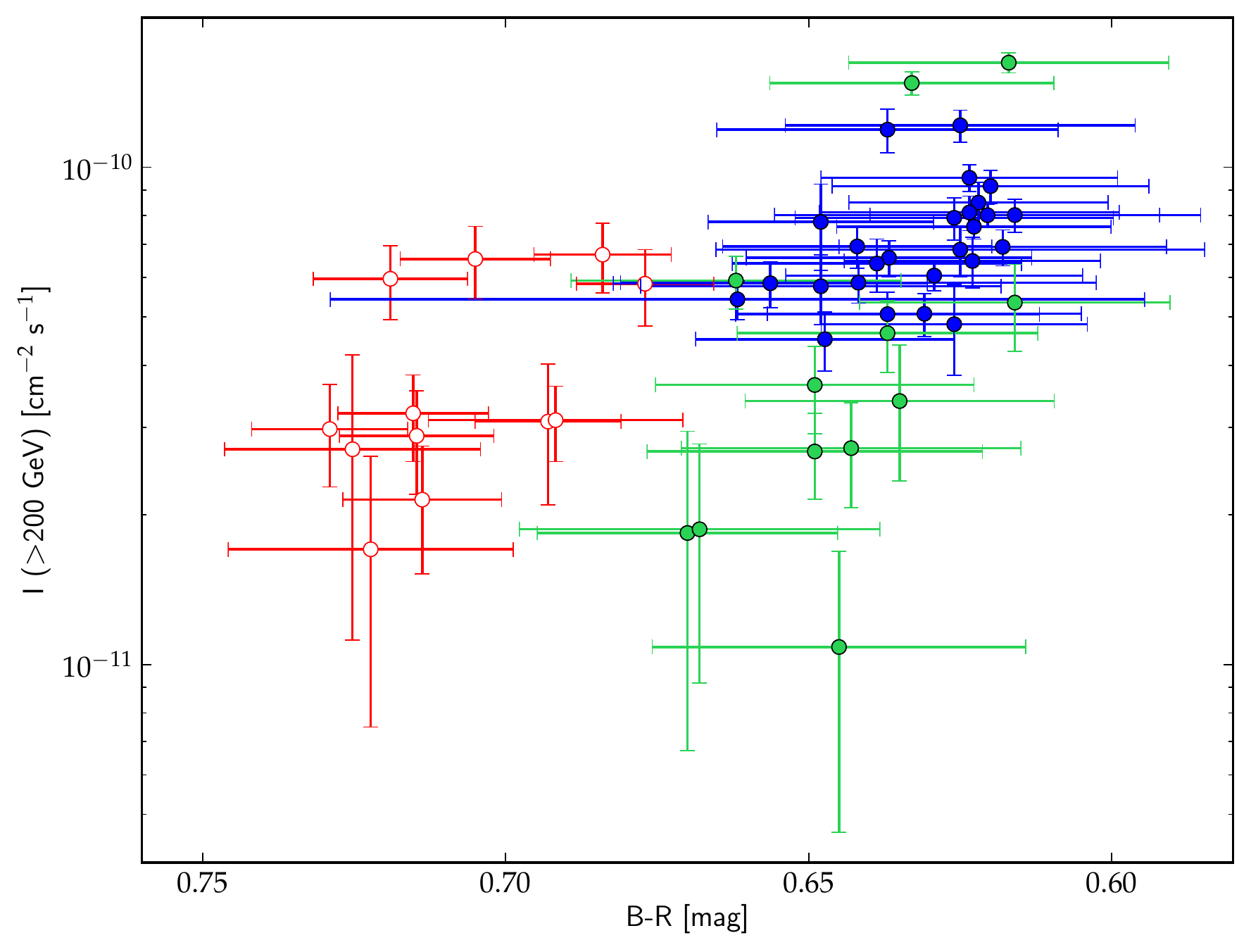}}
\caption[]{VHE $\gamma$-ray photon flux $I_{> 200\,{\rm GeV}}$ vs. $B-R$ colour for \pks\
during the 2007-2009 period. The applied colour coding is the same as in Figures~\ref{atom_lc}
and \ref{hess_b}. }
  \label{hess_bmr}
\end{figure}

Similarly, no strict correlation between the VHE $\gamma$-ray fluxes and optical colours can be found 
for \pks\ during the analysed 2007-2009 period. On the other hand, distinct spectral states of the source
discussed in the previous sections (\S~3.2-3.3) are recovered in the VHE flux vs. $B-R$ colour diagram 
shown in Figure~\ref{hess_bmr}.

\subsection{Broadband SEDs}

\begin{figure}[]
\centering{\includegraphics[width=0.48\textwidth]{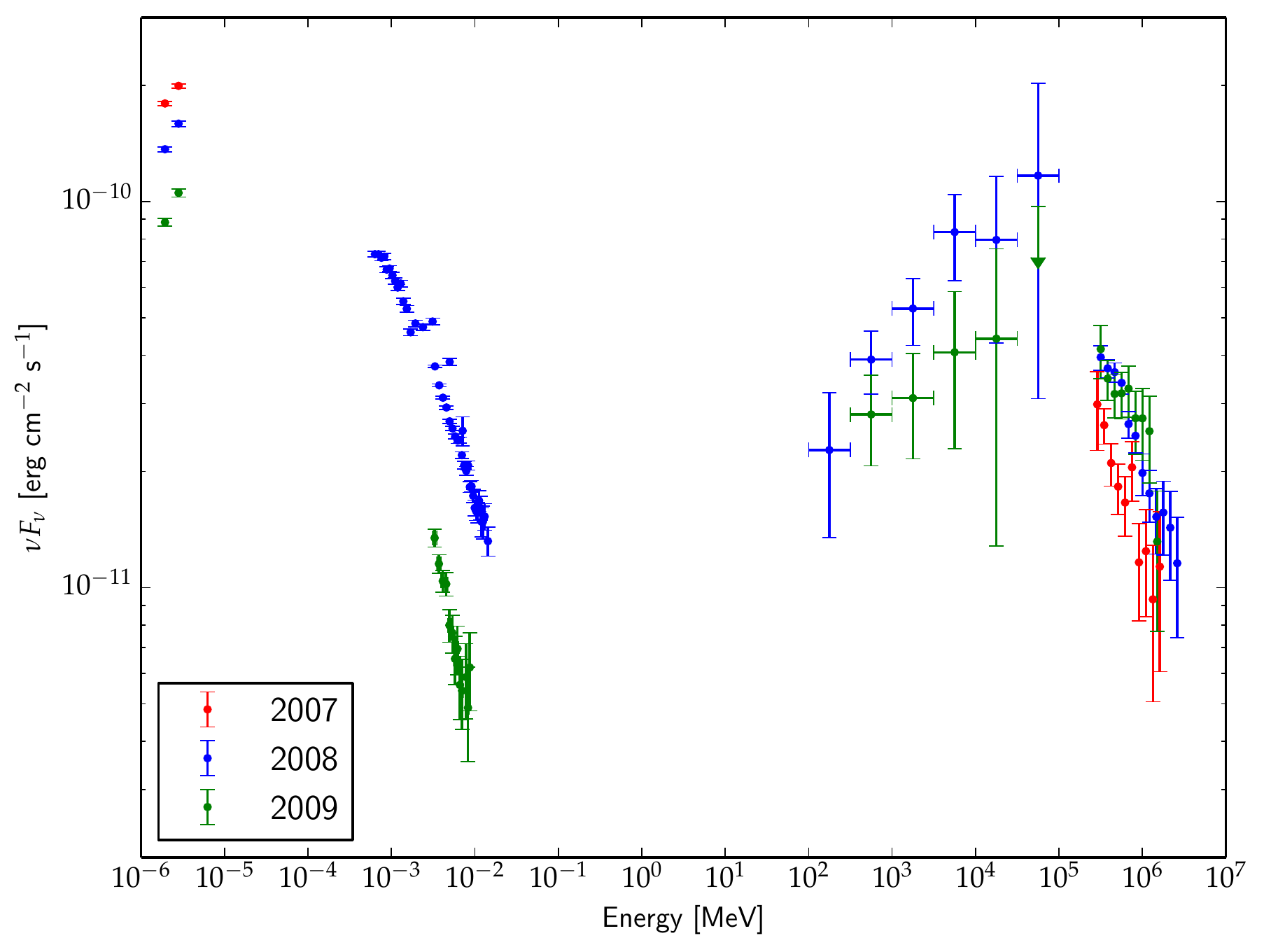}}
\caption[]{Average broadband SEDs of \pks\ for the years 2007, 2008, and 2009 (red, blue and green
symbols, respectively), constructed using quasi-simultaneous data collected with ATOM, RXTE, {\it Swift}, {\it Fermi}-LAT, and H.E.S.S.}
\label{sed}
\end{figure}

Figure~\ref{sed} presents the quasi-simultaneous broadband SEDs of 
\pks\ corresponding to the 2007, 2008, and 2009 epochs (red, blue, and green symbols, respectively).
In the figure we included the VHE $\gamma$-ray data from H.E.S.S. corrected for the absorption 
on the extragalactic background light (EBL) using the EBL model by \cite{EBLmodel}, HE $\gamma$-ray data
from {\it Fermi}-LAT, X-ray data from {\it Swift} and RXTE, and the optical data from ATOM ($B$ and $R$ bands), 
all analysed as described in \S~2 and below, and corresponding to the observation dates provided in Table~\ref{table_5}.

The H.E.S.S. spectra for different epochs considered are reliably fitted by the power-law model 
$dN/dE=N_0 \, (E/E_0)^{-\Gamma}$, where $N_0$ is the normalization of the differential
photon flux and $\Gamma$ is the photon index. The derived model parameters are 
given in  Table~\ref{table_spec} for the fixed photon energy $E_0=1$\,TeV.
The {\it Fermi}-LAT spectra were derived here as described in \S~\ref{fermi_data}, and modelled with 
power-law distributions with the normalization $({5.08\pm 0.54}) \times 10^{-14}$\,MeV$^{-1}$\,s$^{-1}$\,cm$^{-2}$ 
(for $E_0 = 1$\,GeV) and the photon index $\Gamma_{\rm HE} =1.75\pm0.08$ in 2008, and 
$({2.63\pm 0.43}) \times 10^{-14}$\,MeV$^{-1}$\,s$^{-1}$\,cm$^{-2}$ and $\Gamma_{\rm HE} =1.88\pm0.13$ in 2009.
The constant fit to HE light curve during the two analysed epochs returns $\chi^2/\textit{ndof}=9.17/5$ in 2008 and $\chi^2/\textit{ndof}=4.0/6$ in 2009.
The average RXTE/{\it Swift} spectra of \pks\ for the 2008 epoch 
are taken from \cite{varLAT}. The 2009 X-ray spectrum was derived from the 
archival RXTE data (see \S~\ref{rxte_data}), and fitted with a power-law model 
with a normalization of ${3.57^{+0.98}_{-0.75}} \times 10^{-2}$\,keV$^{-1}$\,s$^{-1}$\,cm$^{-2}$ 
and the X-ray photon index $\Gamma_{\rm X} =3.21\pm0.16$.

Figure~\ref{sed} shows that the year-averaged broadband spectra of \pks\
during the analysed period 2007-2009 are characterized by only moderate flux changes at optical, 
X-ray, HE, and VHE $\gamma$-ray frequencies. However, while a systematic spectral 
hardening accompanied by the decrease in the total flux can be seen in the optical range between 
2008 and 2009, the X-ray spectrum steepens during the same time interval (at least within 
the RXTE band). In the HE and VHE $\gamma$-ray regimes the uncertainties in the derived values of the 
photon indices and fluxes are relatively large, precluding therefore any definitive statements on
spectral changes, although it seems that these are very minor, if present at all. The 2007$^{(1)}$ epoch
remains distinct in the SED representation, because it is characterized by the highest flux and the softest
spectrum in the optical range, but at the same time the lowest flux in the VHE range. Unfortunately,
no simultaneous X-ray or HE $\gamma$-ray data exist for this epoch.

Another point to notice in Figure~\ref{sed} is that the overall shape of the `two-bump' optical--to--$\gamma$-ray 
continuum suggests the peak of the synchrotron spectral component is located around optical/UV frequencies, 
while the peak of the IC component is located in the HE range. In addition, the synchrotron peak luminosities are very roughly 
comparable to (but still higher than) the IC peak luminosities for every year considered, so that the bulk of the 
radiatively dissipated power of the source is emitted at optical/UV frequencies. These are the typical characteristics 
of an HBL source in its average/quiescent state in general. The caveat to the above comments is a very likely
multicomponent character of the broadband spectrum  of \pks\ \citep[see][and Sect.,4 below]{var4}.

\begin{table*} 
\centering
\begin{tabular}{c|c|c|c|c|c|c}
\hline
\hline
 Year & Month &  H.E.S.S. & {\it Fermi}-LAT & RXTE & {\it Swift}  & ATOM \\
\hline
2007& June &  14-20  & -- &  -- & --& 20 \\
 & July &  12-13,15-16, 18-21, 23   & -- &  -- & -- & 13,15-16, 18,21, 23  \\
  & August & 6-7, 9-11, 13-16, 19    & -- &  -- & -- & 6-7, 9-11, 13-16, 19  \\
 & October &  1-2  & -- &  -- & -- & 1-2 \\
\hline
2008 &  June & 3-7, 29-30 & -- &  -- & --& 5,6, 29-30 \\
 & July  & 1-5, 26-31   & -- &  -- & --& 1-4, 26-28   \\
 & August & 1-4, 22-31   & 22-31 &  25-31 & 25-31& 1,2,4, 23-31    \\
  & September & 1-6   & 1-6 &  1-5 & 1-5 & 1-6   \\
\hline
2009 & June & 24   & 24 &  -- & --& 24 \\ 
 & July & 19-29, 31 & 19-29, 31 &  28, 31 & -- & 19-29, 31 \\
\hline
\end{tabular}
\caption[]{Dates of the MWL observations of \pks\ during the 2007-2009 period included in the broadband SED
presented in Figure~\ref{sed}.}
  \label{table_5}
\end{table*}
 
\section{Discussion and conclusions}

After the pioneering detections of the brightest HBLs with the previous-generation
Cherenkov telescopes,  the X-ray/VHE $\gamma$-ray connection  was widely 
studied and analysed in the context of multiwavelength variability of BL Lac objects. The first extensive
campaigns revealed that the X-ray and VHE $\gamma$-ray flux changes
are closely related, although the exact correlation patterns emerging from the collected datasets
seemed to vary from object to object and from epoch to epoch even for the same source 
\citep[e.g.][]{krawczynski04,blazejowski05,fossati08}. These results regarded, however, almost 
exclusively flaring states of the brightest objects only, as limited sensitivity of the previously
available instruments precluded  the detection of the dimmer sources or
the bright ones in their quiescent states. The situation has changed recently thanks
to the operation of the modern-generation Cherenkov telescopes (H.E.S.S., MAGIC, VERITAS),
which enabled frequent and less biased (with respect to the flaring activity) observations of bright HBLs, 
as well as the detections of the dimmer (in the VHE range) blazars in particular `low-frequency peaked' 
BL Lac objects (LBLs) \citep[e.g.][]{BLLacMagic}. These new observations revealed or at least 
suggested positive correlations between the optical and VHE $\gamma$-ray frequencies for many
objects, and indeed, the broadband modelling of TeV BL Lac objects in their quiescent states,
like  Mrk 501 and Mrk 421 \citep[respectively]{LAT501,LAT421} or \pks\ itself \citep{varLAT},
 is consistent with the idea that these are electrons emitting synchrotron photons at optical frequencies  
which produce the bulk of the observed $\gamma$ rays via the SSC process.

The ATOM and H.E.S.S. observations presented and discussed in this paper constitute the
most detailed up-to-date insight into the optical--VHE $\gamma$-ray connection for the HBL-type
blazar \pks, spanning a period of three years. During that time, the source was  transitioning 
from its flaring to quiescent optical states, and was characterized by only moderate flux changes at 
different wavelengths and a synchrotron dominance (energy flux ratio $[\nu F_{\nu}]_{\rm ssc} / [\nu F_{\nu}]_{\rm syn} <1$). 
 The observed optical--VHE correlations are rather convoluted.
In particular, we did not find any universal relation between the flux changes in both bands 
on the timescales from days and weeks up to years. We did find, however, two distinct states
of the source at higher optical fluxes, characterized by different optical colours and  possibly also
different $\gamma$-ray spectral properties. 

In the context of the results of our analysis, it is important to note that during the exceptional
VHE outbursts of \pks\ in 2006, when ultra-fast variability at $\gamma$-ray frequencies was detected
and the source displayed a significant Compton-dominance $[\nu F_{\nu}]_{\rm ssc} / [\nu F_{\nu}]_{\rm syn} \gg 1$
\citep{var1}, no optical--VHE correlations were present \citep{var2}, with the possible exception of the delayed
and small-amplitude optical echo \citep{var4}. Such states are rare however,
not only in the case of \pks, but also for the other VHE-detected BL Lac objects. And analogously, the states
with high-amplitude ultra-fast optical variability also seem to be restricted to rare isolated epochs 
\citep{Paltani97,Heidt97}. The bulk of the time-averaged 
radiatively dissipated power of BL Lac objects is therefore released during the extended 
periods of a source quiescence, characterized by moderate flux changes and longer variability timescales.
This general picture is confirmed by the long-term monitoring of \pks\ at different frequencies 
revealing that the typical state of the source corresponds to the lower-activity epochs with the characteristic 
variability timescales of the order of a few/several days in X-rays \citep[e.g.][]{Kataoka01}, and 
months in optical \citep{Kastendieck}. This is exactly the state probed by the observations presented 
and analysed in this paper.

\begin{table*} 
\centering
\begin{tabular}{c|c|c|c|c|c|c}
\hline
\hline
 Year & $\Gamma_{\rm VHE}$  & $N_{\rm 0,\,VHE}$  & $\Gamma_{\rm HE}$  & $N_{\rm 0,\,HE}$ & $\Gamma_{\rm X}$  & $N_{\rm 0,\,X}$\\
    (1)  & (2)  & (3)  & (4) & (5) & (6) & (7)  \\
\hline
2007 & 3.40 $\pm$ 0.09 & 2.48 $\pm$ 0.09 & -- & -- & -- & --\\
2008 & 3.35 $\pm$ 0.04 & 4.30 $\pm$ 0.07 & 1.75 $\pm$ 0.08 & 5.08
$\pm$ 0.54 & -- & --\\
2009 & 3.12 $\pm$ 0.06 & 5.20 $\pm$ 0.18  & 1.88 $\pm$ 0.13 & 2.63 $\pm$ 0.43 & 3.21 $\pm$ 0.16 & $3.57^{+0.98}_{-0.75}$\\
\hline
\end{tabular}
\caption[]{The derived averaged spectral parameters for \pks\ in different epochs discussed in this paper: (1) year of the observation; (2) VHE photon index; (3) normalization of the VHE photon flux at 1\,TeV in units of $10^{-10}$\,TeV$^{-1}$\,s$^{-1}$\,cm$^{-2}$; (4) HE photon index;  (5) normalization of the HE photon flux at 1\,GeV in  units of $10^{-14}$\,MeV$^{-1}$\,s$^{-1}$\,cm$^{-2}$; (6) X-ray photon index; (7) normalization of the X-ray photon flux at 1\,keV in  units of $10^{-2}$\,keV$^{-1}$\,s$^{-1}$\,cm$^{-2}$. The broken power-law model fits to the X-ray data gives photon indices $\Gamma_1$=2.36$\pm$0.01, $\Gamma_2$=2.6$\pm$0.01, and break energy E$_{br}$=4.44$\pm$0.48\,keV \cite{varLAT}.}
  \label{table_spec}
\end{table*}

What we find, however, is that even the average state of \pks\ is characterized by complex multiwavelength 
correlation patterns. Some general trends observed could be explained, at least qualitatively, assuming a modulation 
in the internal parameters of a single dominant emission zone within the outflow. Higher magnetization, for example, 
would correspond to the state with the elevated synchrotron luminosity, steeper optical spectrum due to the 
increased synchrotron cooling rate, but little or no increase in the observed $\gamma$-ray flux, as observed 
during the first half of 2007. At the same time, a more detailed insight into the data, revealing that the optical and 
VHE $\gamma$-ray fluxes are correlated for some epochs but uncorrelated or even 
anti-correlated for the other epochs,  may indicate that this dominant emission zone is not
a single uniform entity. Instead, it may consist of a superposition of various distinct emission components,
characterized by a wider range of the internal parameters.
The reason for the formation of such components is uncertain, but relativistic reconnection events and MHD instabilities
enabling an enhanced energy dissipation, and generating particle beams or compact blobs moving relativistically 
within the outflow, are the widely invoked possibilities, particularly in the context of ultra-rapid $\gamma$-ray
variability of blazar sources \citep[see e.g.][]{bou08,len08,var4}.

\section{Summary}

In this paper we presented the results of the analysis of all the available optical and VHE $\gamma$-ray data for \pks\ 
collected simultaneously with the ATOM and H.E.S.S. telescopes from 2007 until 2009. The gathered data 
constitute the most detailed and up-to-date insights into the optical--VHE $\gamma$-ray connection for this
blazar over the period of three years. During the analysed period, the source was  transitioning from its 
flaring to quiescent optical states, and was characterized by only moderate flux changes at different wavelengths 
and a `bluer--when--brighter' optical behaviour pronounced at higher flux levels. 
There is, however, no universal relation between the VHE $\gamma$-ray and the optical flux changes 
on the timescales from days and weeks to years, as only during the particular epochs is a positive 
correlation seen. At higher optical flux levels the source can enter a distinct state characterized 
by a steep optical spectrum and a low $\gamma$-ray activity.
We argue that the obtained results suggest a complex structure of the dominant emission zone in the \pks\
jet during the quiescent state, possibly by analogy to the multicomponent emission models
discussed before in the context of the spectacular $\gamma$-ray
outbursts of the source in July 2006.

\begin{acknowledgements}
The support of the Namibian authorities and of the University of Namibia
in facilitating the construction and operation of H.E.S.S. is gratefully
acknowledged, as is the support by the German Ministry for Education and
Research (BMBF), the Max Planck Society, the French Ministry for Research,
the CNRS-IN2P3 and the Astroparticle Interdisciplinary Programme of the
CNRS, the U.K. Science and Technology Facilities Council (STFC),
the IPNP of the Charles University, the Czech Science Foundation, the Polish 
Ministry of Science and  Higher Education, the South African Department of
Science and Technology and National Research Foundation, and by the
University of Namibia. We appreciate the excellent work of the technical
support staff in Berlin, Durham, Hamburg, Heidelberg, Palaiseau, Paris,
Saclay, and in Namibia in the construction and operation of the
equipment.
A.W. acknowledge support from the National
Science Center (grant No. 2011/03/N/ST9/01867).
\end{acknowledgements}

\bibliographystyle{aa}
\bibliography{references}

\end{document}